# Thermo-visco-plasticity under high strain rates: a micro-inertia driven dynamic flow rule

Md M Rahaman[1], A Pathak[1], D Roy[1,*], J N Reddy[2]

[1]Computational Mechanics Lab, Department of Civil Engineering, Indian Institute of Science, Bangalore 560012, India

[2]Advanced Computational Mechanics Lab, Department of Mechanical Engineering, Texas A&M University, College Station, Texas 77843-3123

*Corresponding author; email: royd@civil.iisc.ernet.in

## Abstract

*Built on the tenets of rational thermodynamics, this article proposes a theory of strain gradient thermo-visco-plasticity for isotropic polycrystalline materials under high strain rates. The effect of micro-inertia, which arises due to dynamically evolving microstructural defects, is brought to bear on the macro-continuum through a micro-force balance. Constitutive modelling of dissipative micro-stresses incorporates relaxation time parameters to account for the time lags of the dissipative fluxes in attaining a steady state. Augmentation of the micro-force balance with constitutive relations for the micro-stresses yields a non-local flow rule that reflects the effect of micro-inertia on the evolution of the plastic strain. A thermodynamically consistent derivation of temperature evolution is provided, thus replacing an empirical route. Numerical implementation of the proposal does not demand a computationally intensive return mapping algorithm. A two dimensional plane strain model of impact between two 4340 steel plates is used to numerically assess the influence that micro-inertial and relaxation time parameters as well as various length scales may have on the macro-continuum response.*

# 1. Introduction:

A number of experimental studies have established that the strength of materials undergoing inhomogeneous plastic flow has intrinsic size dependence (Nix *et al*., 1998; Fleck *et al*., 1994; Ma *et al.*, 1995; Stölken and Evans, 1998; McElhaney *et al*., 1998; Giannakopoulos and Suresh, 1999; Wright *et al*., 2001; Tymiak *et al*., 2001; Swadenera *et al*., 2002; Haque and Saif, 2003; Shrotriyaa *et al*., 2003). As conventional plasticity theories do not admit intrinsic material length scales, several variants of the non-local plasticity theory have been employed to model inhomogeneous plastic flows. The existing non-local theories such as yield-surface based gradient plasticity or those based on micropolar or micromorphic approaches, typically rely on empirical hardening models incorporated within an algebraic or quasi-static flow rule to describe the evolution of plastic strain. They do not explicitly consider the inertial effect of moving microstructural defects, e.g. moving line defects such as dislocations in polycrystalline solids as emphasized in this work, on the macro-continuum response. On the contrary, dislocation based crystal plasticity theories at high deformation rates (Wang *et al.,* 2007) have highlighted the substantive role of inertial effects associated with dislocation motion. Inertial effects have been demonstrated by considering the acceleration of dislocations in the equation of motion with a proper definition of dislocation mass (Hirth and Lothe, 1982; Wang *et al.,* 2007; Hensen *et. al*., 2013). While these methods perform very well for small volumes and limited parameters, prohibitive computational cost and modelling complexity for larger volumes render them inappropriate for continuum level analyses. This motivates the development of a continuum model that explicates the role of micro-inertia in determining the macro-scale dynamic response. Orowan equation relating the average dislocation velocity and equivalent plastic strain rate could be considered in introducing a micro-inertia term in the continuum model which may reflect, in a phenomenological way, the effect of dislocation acceleration on the macro-continuum response.

A weakly non-local model that could serve as a vehicle for such a development is gradient visco-plasticity (Fleck *et al.*, 1994; Fleck and Hutchinson, 1997; Gurtin, 2000, 2002, 2003). This model incorporates size dependence in the constitutive incorporation of plastic strain gradients that appear with intrinsic length scales, typically encoding characteristic distances between defects (dislocations) and/or their characteristic sizes (dislocation cells). In the context of crystalline plasticity, metal plasticity to wit, dislocations are generally considered as primary micro-structural defects. Of these, only statistically stored dislocations (SSDs) are considered in classical visco-plasticity theories, which pertain only to the case of homogeneous plastic flow. In addition to SSDs, geometrically necessary dislocations (GNDs) are introduced in gradient visco-plasticity theories so as to enforce deformation compatibility (e.g. curvature due to bending) arising from inhomogeneous plastic flow (Nye, 1953; Ashby, 1970). In order to consider the effect of GNDs in the macro model, gradient of the equivalent plastic strain is taken as the additional state variable (Aifantis, 1999; Gurtin, 2000, 2002, 2003). Considering the equivalent plastic strain as an independent variable in addition to displacement, Gurtin has introduced a non-local flow rule by augmenting the constitutive relation for micro-stresses with an additional micro-force balance that may be looked upon as a generalization over both Ginzburg-Landau and Cahn-Hilliard equations (Gurtin, 1996).

By proposing a micro-inertia driven flow rule in the form of a modified micro-force balance, we elucidate, in this work, the effect of the micro-inertial force associated with evolving micro-structural defects on the plastic strain. Relaxation time parameters, introduced to allow time lags for the dissipative fluxes before attaining the steady state, are exploited in the constitutive modelling. Supplemented with constitutive relations for the micro-stresses, the micro-force balance leads to a new dynamic flow rule. Alongside, a thermodynamically consistent derivation of temperature evolution is presented, thus circumventing the empirical route usually adopted (employing, for instance, an empirical coefficient for converting plastic work to temperature). Numerical simulations of impact between two 4340 steel plates under high strain rates are undertaken to illustrate how micro-inertia affects the evolving plastic strain and temperature.

## 2. The Dynamic Flow Rule

The theoretical background needed for this development could be found in the work of Gurtin (2000, 2002, and 2003), though the latter is essentially limited to the iso-thermal quasi-static case. The salient features in the constitutive modelling are based on the following considerations, even though the validity of some of these remains questionable in the context of non-equilibrium statistical dynamics of the continuum.

(1) Helmholtz free energy, if it could be defined based on a quasi-equilibrium distribution (see, for instance, Assumption (A5) in Ge and Jiang, 2008), is dependent on gradient of equivalent plastic strain ($\nabla \gamma^p$) and its thermodynamic conjugate force, $\xi^{en}$, represents the energetic component of vector micro-stress. The statistical basis for the existence of this free energy away from equilibrium (see, for instance, equation (23) in Hatano and Sasa 2001) for a definition of free energy in non-equilibrium steady-state) will however not be taken up here.

(2) Within an isotropic plasticity theory (probably implying weak interactions between slip systems at the crystal level), strain hardening is dependent on the equivalent plastic strain ($\gamma^p$) and the effective plastic strain gradient ($\eta^p$). While $\gamma^p$ may be interpreted as the macroscopic quantity associated with attractive non-planar dislocation interactions forming statistically stored dislocations (SSDs), $\eta^p$ is introduced to characterize the enhanced hardening owing to the inhomogeneous intra-granular deformations generating geometrically necessary dislocations (GNDs). $\eta^p$ is thus a scalar measure of Burgers plastic incompatibility tensor.

(3) In order to characterize the entropy generation associated with the motion of GNDs, a dissipative component of vector micro-stress ($\boldsymbol{\xi}^{dis}$) is introduced as the thermodynamically conjugate force to the gradient of the equivalent plastic strain rate.

We presently aim at extending the setup provided by Gurtin and co-workers to the dynamic case that accounts for both the macro- and micro-inertial effects. We will also endeavour to incorporate temperature effects in the constitutive modelling of the dissipative fluxes. Though it should be conceptually straightforward to align the work with the finite deformation theory, our current focus is on the small deformation setup that enables, based on a moving Lagrangian setup, an additive decomposition of the displacement gradient and affords freedom from distinguishing a material descriptor from the spatial.

## 2.1. Kinematics

We present a summary of the kinematics for small deformation plasticity; for a more detailed exposition, see (Gurtin, 2000, 2002, 2003).

### 2.1.1. Basic Kinematics

The theory of plasticity for solids undergoing small deformation assumes the following decomposition of the displacement gradient field.

$$\nabla \mathbf{u} = \mathbf{H}^e + \mathbf{H}^p \tag{2.1}$$

$\mathbf{u}(\mathbf{x},t)$ denotes the displacement of an arbitrary point $\mathbf{x}$ in the domain of space occupied by the body $\mathcal{B}$. $\mathbf{H}^e$ and $\mathbf{H}^p$ are the elastic and plastic distortion tensors respectively. Elastic and plastic strains are respectively defined using the symmetric tensors:

$$\mathbf{E}^e = \frac{1}{2}\left(\mathbf{H}^e + \mathbf{H}^{eT}\right) \tag{2.2}$$

$$\mathbf{E}^p = \frac{1}{2}\left(\mathbf{H}^p + \mathbf{H}^{pT}\right) \tag{2.3}$$

$\mathbf{E}^p$ is assumed to be deviatoric in line with the oft-used principle of volume invariance under plastic flow, e.g. by dislocation glide or cross-slip (with the latter potentially increasing the number of active slip planes) in polycrystals. Dislocation climb may be ignored under the assumption of plastic incompressibility (Aifantis, 1987). Note that absorbing the plastic part of rotation in its elastic part does not affect the resulting field equations in isotropic plasticity. Accordingly, plastic rotation is considered to be zero with the entire rotation assumed to be elastic. With these assumptions, decomposition of the displacement gradient is given by:

$$\nabla \mathbf{u} = \mathbf{H}^e + \mathbf{E}^p \tag{2.4}$$

The assumption of irrotational plastic flow implies that dissipation due to plastic spin is ignored in this work. However, plastic spin could become thermodynamically important especially for nano-crystalline solids (Wang *et al.*, 2014) as dislocation climb near the grain boundary may lead to grain rotation.

The plastic strain rate ($\dot{\mathbf{E}}^p$) is defined as,

$$\dot{\mathbf{E}}^p = \upsilon^p \mathbf{N}^p, \upsilon^p = \left|\dot{\mathbf{E}}^p\right| \geq 0, \mathbf{N}^p = \frac{\dot{\mathbf{E}}^p}{\left|\dot{\mathbf{E}}^p\right|} \text{ whenever } |\dot{\mathbf{E}}^p| \neq 0 \tag{2.5}$$

where $\upsilon^p$ and $\mathbf{N}^p$ are the equivalent plastic strain rate and the direction of the plastic flow respectively. The equivalent plastic strain is defined as,

$$\gamma^p(x,t) \triangleq \int_0^t \upsilon^p(\mathbf{x},\varsigma)d\varsigma \qquad (2.6)$$

**2.1.2. Burgers Tensor and Equivalent Plastic Strain**

Although the displacement gradient is compatible with a continuously differentiable displacement field, its two components as shown in equation 2.4, are generally not. This may be understood based on an intermediate configuration following elastic unloading in a director based generalized continuum model (see, for instance, Figure 1 in Naghdi and Srinivasa, 1993). The macroscopic Burgers vector acts as a measure of this incompatibility in the plastic strain, subject to the constraint $curl\mathbf{E}^p = -curl\mathbf{H}^e$. A local Burgers vector in continuum form could be defined using Stokes' theorem,

$$b(\Gamma) \triangleq \int_\Gamma \mathbf{E}^p dx = \int_{\partial\Omega} \left(curl\mathbf{E}^p\right)^T \mathbf{n} dA. \qquad (2.7)$$

$\Gamma$ denotes the closed boundary of an arbitrary surface $\partial\Omega$ in the body $\mathcal{B}$ with $\mathbf{n}$ denoting the unit normal field on $\partial\Omega$. Defining the Burgers tensor as

$$\mathbf{G} \triangleq curl\mathbf{E}^p \qquad (2.8)$$

provides a measure of the macroscopic Burgers vector. Rate of the Burgers tensor is written as,

$$\dot{\mathbf{G}} = curl(\upsilon^p \mathbf{N}^p) \qquad (2.9)$$

which can be further expanded as,

$$\dot{\mathbf{G}} = \left(\nabla \upsilon^p \times\right)\mathbf{N}^p + \upsilon^p curl\mathbf{N}^p, \quad \text{where } \left(\nabla \upsilon^p \times\right)_{is} \triangleq \varepsilon_{irs}\upsilon^p_{,r} \quad (2.10)$$

where $\varepsilon_{irs}$ is the alternating tensor. Adopting a simplification in the theory for it to be based on $\nabla \upsilon^p$ only and not on $\nabla \dot{\mathbf{E}}^p$, the term involving $curl\mathbf{N}^p$ (which is the covariant derivative of the orientation tensor and hence a higher order term) is neglected and the rate of the Burgers tensor is thus approximated as,

$$\dot{\mathbf{G}} \approx \left(\nabla \upsilon^p \times\right)\mathbf{N}^p \quad (2.11)$$

The effective plastic strain gradient $\eta^p$ is defined in following manner,

$$\dot{\eta}^p \triangleq \left|\dot{\mathbf{G}}\right|, \text{ and } \eta^p(x,t) \triangleq \int_0^t \dot{\eta}^p(x,\varsigma)d\varsigma \quad (2.12)$$

## 2.2. Balance Laws

Local forms of the conservation laws for mass, momentum and moment of momentum, which hold for every material point, are as follows

$$\dot{\rho} = -\rho \nabla \cdot \mathbf{v} \quad (2.13)$$

$$\rho \dot{\mathbf{v}} = \nabla \cdot \boldsymbol{\sigma} + \mathbf{b} \quad (2.14)$$

$$\boldsymbol{\sigma} = \boldsymbol{\sigma}^T \quad (2.15)$$

where $\rho$ is the mass density, $\mathbf{v}$ the velocity of particle, $\boldsymbol{\sigma}$ the symmetric Cauchy stress tensor and $\mathbf{b}$ is the body force. $\frac{d}{dt} = (\dot{\square})$ is the material time derivative (i.e. derivative taken in moving Lagrangian frame).

A micro-force balance is proposed based on the premise that the "*rate of change of momentum associated with the rate of plastic deformation equals to all forces arising from*

*(and maintaining) the effect of plastic deformation*" (Naghdi and Srinivasa, 1993). For any arbitrary sub-domain $\Omega$ with boundary $\partial\Omega$,

$$\frac{d}{dt}\left[\int_\Omega \rho l_m^2 \upsilon^p d\Omega\right] = \int_{\partial\Omega} \aleph(\mathbf{n})d(\partial\Omega) + \int_\Omega (\tau - \pi)d\Omega \qquad (2.16)$$

Here $\upsilon^p = \dot{\gamma}^p$; $l_m$ is a length scale associated with the (fictive) mass of the micro-structural defects (e.g. dislocation); $\aleph$ is the micro-stress; $\tau$ is the resolved shear stress in the direction of the plastic flow, i.e. $\tau = \boldsymbol{\sigma}:\mathbf{N}^p$; $\pi$ is the micro-stress conjugate to the equivalent plastic strain. In (2.16) micro-inertia term $\rho l_m^2 \upsilon^p$ is introduced in a phenomenological way using the fact that equivalent plastic strain rate ($\upsilon^p$) is related with average dislocation velocity through Orowan equation $\upsilon^p = b\rho_m v_d$ where $b$, $\rho_m$ and $v_d$ are the Burgers vector magnitude, mobile dislocation density and average dislocation velocity, i.e. $v_d = <V_d>$. Here the expectation $<.>$ corresponds to the ensemble averaging of $V_d$, the random dislocation velocity, over a representative volume element (RVE). While a few of the dislocations move very fast (some of them perhaps with speed more than sound) under high strain rates, most of them however move with far lower speed. Thus the distribution of the velocity $V_d$ has a long tail (see, for instance, Figure 8, Wang *et al.*, 2014 and also LeBlanc *et al.*, 2012 for distribution of $V_d$ near depinning transition) so that the average $v_d$ alone is insufficient to accurately describe $V_d$, requiring the incorporation of at least the second moment information. Presently however, we ignore the entropic effects of such fluctuations in $V_d$ and work with only $v_d$. Accordingly, squared length scale ($l_m^2$) associated with the participating mass is also considered a constant. Now, using the second transport theorem and considering $\aleph(\mathbf{n}) = \boldsymbol{\xi}\cdot\mathbf{n}$ in (2.16),

$$\Rightarrow \int_\Omega \rho l_m^2 \ddot{\upsilon}^p d\Omega = \int_{\partial\Omega} \boldsymbol{\xi} \cdot \mathbf{n} \; d(\partial\Omega) + \int_\Omega (\tau - \pi) d\Omega$$

$$\Rightarrow \int_\Omega \rho l_m^2 \ddot{\upsilon}^p d\Omega = \int_\Omega \nabla \cdot \boldsymbol{\xi} \; d\Omega + \int_\Omega (\tau - \pi) d\Omega \quad \text{(using Stokes' theorem)}$$

$$\int_\Omega \rho l_m^2 \ddot{\upsilon}^p d\Omega = \int_\Omega \left[\nabla \cdot \boldsymbol{\xi} + (\tau - \pi)\right] d\Omega \tag{2.17}$$

Applying the localization theorem on (2.17) leads to the following local form for the micro-force balance

$$\rho l_m^2 \ddot{\upsilon}^p = \nabla \cdot \boldsymbol{\xi} + (\tau - \pi) \tag{2.18}$$

Note that the mass associated with the micro-structural defects (e.g. dislocations) is fictitious and must be interpreted in a generalized sense in so far as it participates in the generation of kinetic energy associated with the defect motion (Hirth and Lothe 1992).

## 2.3. Thermodynamics: First Law and Internal Energy Evolution

The subject of non-equilibrium statistical thermodynamics, which should provide the right setup for the modelling of high strain-rate continuum plasticity, is still in its incipient stages (Attard, 2012). Thus, we make use of the rational thermodynamics thereby assuming that such thermodynamic state functions as entropy, internal energy and free energy make sense. The first law of thermodynamics on the conservation of energy could be stated as follows:

$$dE = \delta W + \delta Q \quad (\delta \text{ indicates an inexact differential}) \tag{2.19}$$

In rate form, we may write

$$\frac{dE}{dt} = W^\circ + Q^\circ \tag{2.20}$$

where the total energy $E$ equals the sum of internal energy ($U$) and kinetic energy ($K$), i.e. $E = K + U$. $W$ and $Q$ are respectively the external work and the heat exchanged with the constant temperature heat bath. $W$ and $Q$ are path dependent and hence not state functions.

$(.)^o$ denotes the rate of an inexact differential, signifying path dependence. Internal energy $U$ may be written in terms of the specific internal energy $e$ as,

$$U = \int_\Omega \rho e d\Omega \tag{2.21}$$

Then the rate of change of internal energy is

$$\frac{dU}{dt} = \int_\Omega \rho \dot{e} d\Omega \tag{2.22}$$

Similarly the rate form of the kinetic energy could be written, considering contributions from its conventional macroscopic and the presently introduced microscopic constituents. First we write $K$ in terms of its macro- and micro-components:

$$K = \int_\Omega \left[ \frac{1}{2} \rho \mathbf{v} \cdot \mathbf{v} + \frac{1}{2} \rho l_m^2 \left( \upsilon^p \right)^2 \right] d\Omega \tag{2.23}$$

Rate of change of kinetic energy follows as,

$$\frac{dK}{dt} = \int_\Omega \left[ \rho \dot{\mathbf{v}} \cdot \mathbf{v} + \rho l_m^2 \dot{\upsilon}^p \upsilon^p \right] d\Omega \tag{2.24}$$

Now an expression for the external power may be written considering the macroscopic part corresponding to the traction $\mathbf{t}(\mathbf{n})$ and body force $\mathbf{b}$, and the microscopic part corresponding to the micro-force $\aleph(\mathbf{n})$ acting on the boundary of the domain.

$$\begin{aligned} W^\circ &= \int_{\partial\Omega} \mathbf{t}(\mathbf{n}) \cdot \mathbf{v} \, d(\partial\Omega) + \int_\Omega \mathbf{b} \cdot \mathbf{v} \, d\Omega + \int_{\partial\Omega} \aleph(\mathbf{n}) \upsilon^p d(\partial\Omega) \\ &= \int_{\partial\Omega} \boldsymbol{\sigma} \mathbf{n} \cdot \mathbf{v} d(\partial\Omega) + \int_\Omega \mathbf{b} \cdot \mathbf{v} \, d\Omega + \int_{\partial\Omega} \boldsymbol{\xi} \cdot \mathbf{n} \upsilon^p d(\partial\Omega) \\ &= \int_{\partial\Omega} \left( \boldsymbol{\sigma}^T \mathbf{v} \right) \cdot \mathbf{n} d(\partial\Omega) + \int_\Omega \mathbf{b} \cdot \mathbf{v} \, d\Omega + \int_{\partial\Omega} \boldsymbol{\xi} \upsilon^p \cdot \mathbf{n} d(\partial\Omega) \\ &= \int_\Omega \nabla \cdot \left( \boldsymbol{\sigma}^T \mathbf{v} \right) d\Omega + \int_\Omega \mathbf{b} \cdot \mathbf{v} \, d\Omega + \int_\Omega \nabla \cdot (\boldsymbol{\xi} \upsilon^p) d\Omega \end{aligned}$$

$$W^\circ = \int_\Omega \left[ \{ (\nabla \cdot \boldsymbol{\sigma} + \mathbf{b}) \cdot \mathbf{v} + \boldsymbol{\sigma} : \nabla \mathbf{v} \} + \{ (\nabla \cdot \boldsymbol{\xi}) \upsilon^p + \boldsymbol{\xi} \cdot \nabla \upsilon^p \} \right] d\Omega \tag{2.25}$$

Considering the heat source $h$ and the heat flux vector $\mathbf{q}$, the thermal power expression is obtained as follow:

$$Q^\circ = \int_\Omega \rho h \, d\Omega - \int_{\partial\Omega} \mathbf{q} \cdot \mathbf{n} \, d(\partial\Omega)$$

$$= \int_\Omega \rho h \, d\Omega - \int_\Omega \nabla \cdot \mathbf{q} \, d\Omega$$

$$Q^\circ = \int_\Omega (\rho h - \nabla \cdot \mathbf{q}) d\Omega \tag{2.26}$$

Substituting (2.22), (2.24), (2.25) and (2.26) in (2.20), the first law of thermodynamic takes the following form:

$$\int_\Omega \left[\rho \dot{\mathbf{v}} \cdot \mathbf{v} + \rho l_m^2 \ddot{\upsilon}^p \cdot \upsilon^p + \rho \dot{e}\right] d\Omega = \int_\Omega \left[\{(\nabla \cdot \boldsymbol{\sigma} + \mathbf{b}) \cdot \mathbf{v} + \boldsymbol{\sigma} : \nabla \mathbf{v}\}\right] d\Omega$$
$$+ \int_\Omega \left[\{(\nabla \cdot \boldsymbol{\xi})\upsilon^p + \boldsymbol{\xi} \cdot \nabla \upsilon^p\} + \rho h - \nabla \cdot \mathbf{q}\right] d\Omega \tag{2.27}$$

Further, using conservation of linear momentum (2.14) and conservation of angular momentum (2.18) in (2.27), we obtain the following equation

$$\int_\Omega \rho \dot{e} d\Omega = \int_\Omega \left[\boldsymbol{\sigma} : \nabla \mathbf{v} + (\pi - \tau)\upsilon^p + \boldsymbol{\xi} \cdot \nabla \upsilon^p + \rho h - \nabla \cdot \mathbf{q}\right] d\Omega \tag{2.28}$$

An evolution equation for the specific internal energy results upon localization of (2.28) which can be written as

$$\rho \dot{e} = \boldsymbol{\sigma} : \nabla \mathbf{v} + (\pi - \tau)\upsilon^p + \boldsymbol{\xi} \cdot \nabla \upsilon^p + \rho h - \nabla \cdot \mathbf{q} \tag{2.29}$$

## 2.4. Constitutive Modelling

In this section we present the constitutive equations obtained upon an imposition of the second law of thermodynamics. In this theory, the Helmholtz free energy ($\Psi$) is assumed to depend constitutively on the elastic strain, the gradient of the equivalent plastic strain and temperature $T$. Owing to the presence of the configurational defect elements such as dislocations, interstitials and point defects etc., associating $T$ with the microscopic kinetic-vibrational degrees of freedom (e.g. atomic vibration) alone may not be correct. To address this issue, Langer and co-workers (Langer *et al.*, 2010; Langer, 2015) employ two distinct thermodynamic fragments, viz. the configurational and kinetic-vibrational subsystems, such that the so-called effective temperature in the cofigurational subsystem is much higher than the kinetic-vibrational one. Moreover, the heat flow across the two subsystems provides yet another source of non-locality. For the present, however, we avoid such nuances and work with a single intensive temperature variable *T*. With this, the free energy is expressed as

$$\Psi = \Psi(\mathbf{E}^e, \nabla \gamma^p, T) \tag{2.30}$$

The differential of the free energy, a state function, is then written as,

$$d\Psi = \frac{\partial \Psi}{\partial \mathbf{E}^e} : d\mathbf{E}^e + \frac{\partial \Psi}{\partial \nabla \gamma^p} \cdot d(\nabla \gamma^p) + \frac{\partial \Psi}{\partial T} dT \tag{2.31}$$

From (2.31), we can also write the time derivative of Helmholtz free energy as,

$$\dot{\Psi} = \frac{\partial \Psi}{\partial \mathbf{E}^e} : \dot{\mathbf{E}}^e + \frac{\partial \Psi}{\partial \nabla \gamma^p} : \nabla \dot{\gamma}^p + \frac{\partial \Psi}{\partial T} \dot{T} \tag{2.32}$$

A Legendre transform shifting the dependence from *s* to *T* enables writing the Helmholtz free energy in terms of the internal energy as follow:

$$\Psi = e - Ts \tag{2.33}$$

where *s* is the specific entropy. Appealing to (2.33), one can rearrange the equation above as:

$$\rho T \dot{s} = \rho \dot{e} - \rho \dot{\Psi} - \rho s \dot{T} \tag{2.34}$$

Substituting the expression for the evolution of specific internal energy (2.29) and time derivative of Helmholtz free energy (2.32) in (2.34), we have,

$$\begin{aligned}\rho T \dot{s} = &\boldsymbol{\sigma} : \nabla \mathbf{v} + (\pi - \tau) \upsilon^p + \boldsymbol{\xi} \cdot \nabla \upsilon^p + \rho h - \nabla \cdot \mathbf{q} \\ &- \rho \frac{\partial \Psi}{\partial \mathbf{E}^e} : \dot{\mathbf{E}}^e - \rho \frac{\partial \Psi}{\partial \nabla \gamma^p} : \nabla \dot{\gamma}^p - \rho \frac{\partial \Psi}{\partial T} \dot{T} - \rho s \dot{T}\end{aligned} \tag{2.35}$$

Using $\boldsymbol{\sigma}:\nabla \mathbf{v} = \boldsymbol{\sigma}:\frac{1}{2}(\nabla \mathbf{v}+\nabla \mathbf{v}^T) = \boldsymbol{\sigma}:\dot{\mathbf{E}} = \boldsymbol{\sigma}:\dot{\mathbf{E}}^e + \boldsymbol{\sigma}:\dot{\mathbf{E}}^p = \boldsymbol{\sigma}:\dot{\mathbf{E}}^e + \boldsymbol{\sigma}:\upsilon^p \mathbf{N}^p = \boldsymbol{\sigma}:\dot{\mathbf{E}}^e + \tau \upsilon^p$

and decomposing the energetic and dissipative parts of the vector microstress as $\boldsymbol{\xi} = \boldsymbol{\xi}^{en} + \boldsymbol{\xi}^{dis}$.

we have

$$\rho \dot{s} = \frac{1}{T}\left[\left(\boldsymbol{\sigma} - \rho\frac{\partial \Psi}{\partial \mathbf{E}^e}\right):\dot{\mathbf{E}}^e + \left(\boldsymbol{\xi}^{en} - \rho\frac{\partial \Psi}{\partial \nabla \gamma^p}\right):\nabla \dot{\gamma}^p - \left(\rho\frac{\partial \Psi}{\partial T} + \rho s\right)\dot{T} + \boldsymbol{\xi}^{dis}\cdot \nabla \dot{\gamma}^p + \pi \upsilon^p + \rho h - \nabla \cdot \mathbf{q}\right] \quad (2.36)$$

Now the entropy balance equation is:

$$\rho \dot{s} + \nabla \cdot \mathbf{J}^s = \eta^s + \frac{\rho h}{T} \quad (2.37)$$

where $\eta^s$ denotes the local entropy production rate and $\mathbf{J}^s = \frac{\mathbf{q}}{T}$. Using equation (2.36) in (2.37), we arrive at the following expression for the local rate of entropy production:

$$\eta^s = \frac{1}{T}\left[\left(\boldsymbol{\sigma} - \rho\frac{\partial \Psi}{\partial \mathbf{E}^e}\right):\dot{\mathbf{E}}^e + \left(\boldsymbol{\xi}^{en} - \rho\frac{\partial \Psi}{\partial \nabla \gamma^p}\right):\nabla \dot{\gamma}^p - \left(\rho\frac{\partial \Psi}{\partial T} + \rho s\right)\dot{T} + \boldsymbol{\xi}^{dis}\cdot \nabla \upsilon^p + \pi \upsilon^p\right] - \frac{1}{T^2}\mathbf{q}\cdot \nabla T \quad (2.38)$$

One way of ensuring a non-negative entropy production rate (i.e. $\eta^s \geq 0$) is as follows:

$$\boldsymbol{\sigma} = \rho\frac{\partial \Psi}{\partial \mathbf{E}^e}, \quad \boldsymbol{\xi}^{en} = \rho\frac{\partial \Psi}{\partial \nabla \gamma^p} \quad (2.39)$$

$$\rho s = -\rho\frac{\partial \Psi}{\partial T} \quad (2.40)$$

$$\pi = \mathscr{G}_1\left(\gamma^p, \eta^p, \upsilon^p, |\nabla \upsilon^p|, T\right) \quad (2.41)$$

$$\boldsymbol{\xi}^{dis} = \mathscr{G}_2\left(\gamma^p, \eta^p, \upsilon^p, |\nabla \upsilon^p|, T\right) l_3^2 \nabla \upsilon^p \quad (2.42)$$

$$\mathbf{q} = -\kappa \nabla T \quad (2.43)$$

where, $\mathscr{G}_1, \mathscr{G}_2 \geq 0$ and the heat conductivity $\kappa \geq 0$. Moreover,

$$\boldsymbol{\sigma} = \boldsymbol{C}\mathbf{E}^e, \quad \boldsymbol{\xi}^{en} = S_0 l_1^2 \nabla \gamma^p \quad (2.44)$$

where $\boldsymbol{C}$, $S_0$ and $l_1$ are respectively a fourth order constitutive tensor, initial yield strength of the material and an energetic length scale related to the gradient of the equivalent plastic strain. It is to be noted that equations (2.41)-(2.43) conform to Fourier-type law, leading to infinite speed of propagation of disturbances (Lebon *et al.*, 2008). Considerations of time lag for the dissipative fluxes to attain the steady state condition would yield Maxwell-Cattaneo type equations and thus overcome the paradox of infinite speed of propagation. We thus introduce the time lag relaxation parameters $t_r^\pi$, $t_r^\xi$ and $t_r^q$ for the dissipative fluxes $\pi$, $\xi^{dis}$ and $\mathbf{q}$ respectively and assume the following constitutive equations:

$$\pi(t+t_r^\pi) \approx t_r^\pi \dot{\pi} + \pi = \mathscr{G}_1\left(\gamma^p, \eta^p, \upsilon^p, |\nabla \upsilon^p|, T\right) \tag{2.45}$$

$$\xi^{dis}(t+t_r^\xi) \approx t_r^\xi \dot{\xi}^{dis} + \xi^{dis} = \mathscr{G}_2\left(\gamma^p, \eta^p, \upsilon^p, |\nabla \upsilon^p|, T\right) l_3^2 \nabla \upsilon^p \tag{2.46}$$

$$\mathbf{q}(t+t_r^q) \approx t_r^q \dot{\mathbf{q}} + \mathbf{q} = -\kappa \nabla T \tag{2.47}$$

where,

$$\mathscr{G}_1\left(\gamma^p, \eta^p, \upsilon^p, |\nabla \upsilon^p|, T\right) = S_{grad}(\gamma^p, \eta^p) R_1(d^p) R_2(T) \tag{2.48}$$

$$\mathscr{G}_2\left(\gamma^p, \eta^p, \upsilon^p, |\nabla \upsilon^p|, T\right) = S_{grad} R_1(d^p) R_2(T) \tag{2.49}$$

$$d^p = \sqrt{\left(\upsilon^p\right)^2 + l_3^2 \left|\nabla \upsilon^p\right|^2} \tag{2.50}$$

$$S_{grad}(\gamma^p, \eta^p) = S_0 \sqrt{\left(f\left(\gamma^p\right)\right)^2 + l_2 \eta^p} \tag{2.51}$$

$$f\left(\gamma^p\right) = 1 + \frac{H_0}{S_0}\left(\gamma^p\right)^n \tag{2.52}$$

$H_0$, $l_2$ and $n$ are respectively hardening modulus, dissipative length scale associated with the effective plastic strain and a sensitivity parameter related to strain hardening.

$$R_1(d^p) = \left(\frac{d^p}{d_0}\right)^{m_1} \tag{2.53}$$

$$R_2(T) = 1 - \left(T^*\right)^{m_2} \quad \text{with } T^* = \left(\frac{T - T_r}{T_m - T_r}\right) \tag{2.54}$$

$d_0$, $T_r$, $T_m$, $m_1$ and $m_2$ are respectively the reference strain rate, reference temperature, melting temperature, strain rate hardening sensitivity parameter and temperature softening sensitivity parameter.

## 2.5. Temperature Evolution

We derive the temperature evolution employing the thermodynamic relations and assumed constitutive equations discussed in the sections above. In view of the expression for the differential form of Helmholtz free energy (2.31), restrictions on the forms of Cauchy stress and vector micro stress (2.39) along with their assumed constitution (2.44), we have,

$$d\Psi = \frac{1}{\rho}\boldsymbol{C}\mathbf{E}^e : d\mathbf{E}^e + \frac{1}{\rho}S_0 l_1^2 \nabla\gamma^p \cdot d(\nabla\gamma^p) - sdT$$

$$d\Psi = \frac{1}{\rho}d\left[\frac{1}{2}\boldsymbol{C}\mathbf{E}^e : \mathbf{E}^e + \frac{1}{2}S_0 l_1^2 \nabla\gamma^p \cdot \nabla\gamma^p\right] - sdT \qquad (2.55)$$

Integrating both sides of equation (2.55) leads to,

$$\Psi = \frac{1}{\rho}\left[\frac{1}{2}\boldsymbol{C}\mathbf{E}^e : \mathbf{E}^e + \frac{1}{2}S_0 l_1^2 \nabla\gamma^p \cdot \nabla\gamma^p\right] + \Psi_1(\mathbf{E}^e, \nabla\gamma^p, T) \qquad (2.56)$$

where the function $\Psi_1$ is to be determined. Partial differentiation of $\Psi$ with respect to $\mathbf{E}^e$, $\nabla\gamma^p$ and $T$ lead to the following relations:

$$\frac{\partial \Psi}{\partial E^e} = \frac{1}{\rho}\boldsymbol{C}\mathbf{E}^e + \frac{\partial \Psi_1}{\partial \mathbf{E}^e} \qquad (2.57)$$

$$\frac{\partial \Psi}{\partial \nabla\gamma^p} = \frac{1}{\rho}S_0 l_1^2 \nabla\gamma^p + \frac{\partial \Psi_1}{\partial \nabla\gamma^p} \qquad (2.58)$$

Comparing, from (2.39) and (2.44), descriptions of Cauchy stress with (2.57) and description of energetic part of vector micro-stress with (2.58) we have,

$$\frac{\partial \Psi_1}{\partial E^e} = \frac{\partial \Psi_1}{\partial \nabla\gamma^p} = 0 \qquad (2.59)$$

Therefore, $\Psi_1$ is a function of temperature $T$ only.

$$\text{Say, } \Psi_1 = \Psi_1(T) = \frac{c(T)}{\rho} \tag{2.60}$$

Here $\rho$ is introduced to follow the convention. Substituting (2.60) in (2.56) renders the following expression for Helmholtz free energy:

$$\rho\Psi = \left[\frac{1}{2}\boldsymbol{C}\mathbf{E}^e : \mathbf{E}^e + \frac{1}{2}S_0 l_1^2 \nabla\gamma^p \cdot \nabla\gamma^p\right] + c(T) \tag{2.61}$$

Now taking expression of specific entropy form (2.40) and introducing expression of Helmholtz free energy from (2.61) we have,

$$\rho s = -\rho\frac{\partial\Psi}{\partial T} = -\frac{dc}{dT} \tag{2.62}$$

Considering (2.33), (2.61) and (2.62), the internal energy expression can be written as,

$$\rho e = \frac{1}{2}\boldsymbol{C}\mathbf{E}^e : \mathbf{E}^e + \frac{1}{2}S_0 l_1^2 \nabla\gamma^p \cdot \nabla\gamma^p + \left[c(T) - \frac{dc}{dT}T\right] \tag{2.63}$$

In the case of zero strain, consideration of first law of thermodynamics (2.19) with (2.63) gives the following:

$$\rho de = -T\frac{d^2c}{dT^2} = \delta Q = \rho C_v(T)dT \tag{2.64}$$

where $C_v(T)$ is the specific heat. Now, let us consider the last term of (2.63),

$$\left[c(T) - \frac{dc}{dT}T\right]_T = \left[c(T) - \frac{dc}{dT}T\right]_{T_0} + \int_{T_0}^{T}\left(-T\frac{d^2c}{dT^2}\right)dT = \left[c(T) - \frac{dc}{dT}T\right]_{T_0} + \int_{T_0}^{T}(\rho C_v(T))dT \tag{2.65}$$

Employing (2.65) in (2.63) and differentiating it in time leads to the following:

$$\rho\dot{e} = \frac{1}{2}\boldsymbol{C}\mathbf{E}^e : \dot{\mathbf{E}}^e + \frac{1}{2}S_0 l_1^2 \nabla\gamma^p \cdot \nabla\dot{\gamma}^p + \rho C_v(T)\dot{T} + \rho(\nabla\cdot\mathbf{v})e \tag{2.66}$$

Substituting (2.66) in expression for the evolution of specific internal energy obtained earlier in (2.29), an equation for temperature evolution may be obtained.

$$\dot{T} = \frac{1}{\rho C_v(T)}\left[\pi\upsilon^p + \boldsymbol{\xi}^{dis}\cdot\nabla\upsilon^p + \rho h - \nabla\cdot\mathbf{q} - \rho(\nabla\cdot\mathbf{v})e\right] \tag{2.67}$$

In general, introduction of relaxation time parameters does not affect the non-negativity of the entropy production rate. Through numerical simulations carried out as part of this study, it has been observed that the introduction of a very high (possibly unphysical) relaxation time ($\sim 10^3$ sec) may precipitate a negative entropy production rate. However the magnitude of the physical relaxation time parameters is presently determined by the typical time scale of an impact problem which usually is of the order of microseconds, so the negativity of entropy production rate is not a concern.

Another way of resolving this paradox could have been by introducing correction terms associated with additional relaxation parameters in the entropy production rate. Extended irreversible thermodynamics (Jou and Vázquez, 2001; Alvarez and Casas-Vázquez, 2008) is one such framework which provides a relatively more consistent means to introduce the correction term (see, for instance, Alvarez and Casas-Vázquez, 2008, for the evolution of heat flux). Adoption of this approach in this work would have brought in additional complexities in the constitutive modelling of micro-stresses in the form of coupling terms involving the equivalent plastic strain, temperature and rate of equivalent plastic strain and hence is not attempted.

## 3. SPH-based Simulations

The proposed governing equations and constitutive modelling are solved numerically using smooth particle hydrodynamics (SPH) method, a scheme that, unlike its Galerkin-projected mesh-free and finite element counterparts, bypasses computing the weak form thereby affording freedom from costly numerical integration over elements/background cells. Node-wise enforcement of the balance laws also ensures easier parallelizability. Outlined in this section are the essential steps in the implementation of the SPH. Interested readers are referred to (Libersky and Petscheck, 1991; Shaw and Reid, 2009) for a more detailed exposition of the method and its applications to continuum mechanics.

## 3.1. Discretizing the Governing Equations:

The evolution equations of mass, velocity, internal energy and temperature, derived in Section 3, could be written in the indicial notation as follows:

$$\dot{\rho} = -\rho \frac{\partial \mathbf{v}^{\beta}}{\partial \mathbf{x}^{\beta}} \tag{3.1}$$

$$\dot{\mathbf{v}} = \frac{1}{\rho} \frac{\partial \boldsymbol{\sigma}^{\alpha\beta}}{\partial \mathbf{x}^{\beta}} \left( \text{for } \mathbf{b} = 0 \right) \tag{3.2}$$

$$\dot{\upsilon}^{p} = \frac{1}{\rho l_m^2} \left[ \frac{\partial \boldsymbol{\xi}^{\beta}}{\partial \mathbf{x}^{\beta}} + (\tau - \pi) \right] \tag{3.3}$$

$$\dot{e} = \frac{\boldsymbol{\sigma}^{\alpha\beta}}{\rho} \frac{\partial \mathbf{v}^{\alpha}}{\partial \mathbf{x}^{\beta}} + \frac{(\pi - \tau)\upsilon^{p}}{\rho} + \frac{\boldsymbol{\xi}^{\beta}}{\rho} \frac{\partial \upsilon^{p}}{\partial \mathbf{x}^{\beta}} - \frac{1}{\rho} \frac{\partial \mathbf{q}^{\beta}}{\partial \mathbf{x}^{\beta}} \quad (\text{for } h = 0) \tag{3.4}$$

$$\dot{T} = \frac{1}{C_v(T)} \left[ \frac{\pi \upsilon^p}{\rho} + \frac{\left(\boldsymbol{\xi}^{\text{dis}}\right)^{\beta}}{\rho} \frac{\partial \upsilon^p}{\partial \mathbf{x}^{\beta}} - \frac{1}{\rho} \frac{\partial \mathbf{q}^{\beta}}{\partial \mathbf{x}^{\beta}} - e \frac{\partial \mathbf{v}^{\beta}}{\partial \mathbf{x}^{\beta}} \right] \tag{3.5}$$

where, for any material point $\mathbf{x}^{\beta}$, $\mathbf{v}^{\beta}$, $\boldsymbol{\xi}^{\beta}$, $\left(\boldsymbol{\xi}^{\text{dis}}\right)^{\beta}$, $\mathbf{q}^{\beta}$ and $\boldsymbol{\sigma}^{\alpha\beta}$ are respectively the elements of spatial Cartesian co-ordinate, velocity vector, vector micro-stress, dissipative part of the vector micro-stress, heat flux vector and Cauchy stress tensor with tension taken as positive. The three spatial directions are denoted using the superscripts $\alpha$, $\beta = 1,2,3$. The numerical solution scheme for the governing equations is obtained after converting (3.1)-(3.5) to the following discretized forms.

$$\dot{\rho}_i = \rho_i \sum_j \frac{m_j}{\rho_j} \left( \mathbf{v}_i^{\beta} - \mathbf{v}_j^{\beta} \right) w_{ij,\beta} \tag{3.6}$$

$$\dot{\mathbf{v}}_i^\alpha = \sum_j m_j \left( \frac{\sigma_i^{\alpha\beta}}{\rho_i^2} + \frac{\sigma_j^{\alpha\beta}}{\rho_j^2} + \Pi_{ij} \right) w_{ij,\beta} \tag{3.7}$$

$$\dot{\upsilon}_i^p = \frac{1}{l_m^2} \left[ \sum_j m_j \left( \frac{\xi_i^\beta}{\rho_i^2} + \frac{\xi_j^\beta}{\rho_j^2} \right) w_{ij,\beta} + (\tau_i - \pi_i) \right] \tag{3.8}$$

$$\dot{e}_i = \sum_j m_j \left[ (\mathbf{v}_i^\alpha - \mathbf{v}_j^\alpha)\left( \frac{\sigma_i^{\alpha\beta}}{\rho_i^2} + \frac{1}{2}\Pi_{ij} \right) + (\upsilon_i^p - \upsilon_j^p)\left( \frac{\xi_i^\beta}{\rho_i^2} \right) - (\mathbf{q}_i^\beta - \mathbf{q}_j^\beta)\left( \frac{1}{\rho_i^2} \right) \right] w_{ij,\beta} + \frac{(\pi_i - \tau_i)\upsilon_i^p}{\rho_i} \tag{3.9}$$

$$\dot{T}_i = \frac{1}{(C_v(T))_i} \left[ \sum_j m_j \left\{ (\upsilon_i^p - \upsilon_j^p)\left( \frac{(\xi^{\text{dis}})_i^\beta}{\rho_i^2} \right) - (\mathbf{q}_i^\beta - \mathbf{q}_j^\beta)\left( \frac{1}{\rho_i^2} \right) \right\} w_{ij,\beta} + \frac{\pi_i \upsilon_i^p}{\rho_i} - e_i \sum_j \frac{m_j}{\rho_j}(\mathbf{v}_i^\beta - \mathbf{v}_j^\beta)w_{ij,\beta} \right] \tag{3.10}$$

where, $\mathbf{v}_{ij} = \mathbf{v}_i - \mathbf{v}_j$ and $\mathbf{x}_{ij} = \mathbf{x}_i - \mathbf{x}_j$. $w_{ij} = w(\mathbf{x}_i - \mathbf{x}_j, r_{ij})$ is the kernel function with smoothing length $r_{ij} = (r_i + r_j)/2$, where $r_i$ denotes the smoothing length associated with the $i^{th}$ node. The following cubic kernel function is used.

$$w(\lambda, r) = a_0 \begin{cases} 1 - \frac{3}{2}\lambda^2 + \frac{3}{4}\lambda^3 & 0 \leq \lambda \leq 1 \\ \frac{1}{4}(2-\lambda)^3 & 1 \leq \lambda \leq 2 \\ 0 & \lambda \geq 2 \end{cases} \tag{3.11}$$

where $r$ is the support size of $w$ and $a_0 = \frac{2}{3r}$ in 1D, $\frac{10}{7\pi r^2}$ in 2D and $\frac{1}{\pi r^3}$ in 3D.

To stabilize the SPH computation in the presence of shocks, artificial viscosity ($\Pi_{ij}$) is conventionally used in (3.7) and (3.9). In this work, an acceleration corrected form of $\Pi_{ij}$ is employed (Shaw and Reid, 2009; Shaw et al., 2011; Shaw and Roy, 2012).

$$\Pi_{ij} = \frac{-\alpha_1 C_{ij}\omega_{ij} + \alpha_2 \omega_{ij}^2}{\rho_{ij}} \text{ if } \mathbf{v}_{ij} \cdot \mathbf{x}_{ij} < 0 \text{ else } \Pi_{ij} = 0$$

$$\text{with } \omega_{ij} = \frac{r(\mathbf{v}_{ij} \cdot \mathbf{x}_{ij})}{|\mathbf{x}_{ij}|^2 + \varepsilon r^2}, \quad C_{ij} = \frac{C_i + C_j}{2}, \quad \rho_{ij} = \frac{\rho_i + \rho_j}{2}$$

where, $\alpha_1, \alpha_2$ are artificial viscosity parameters; $\varepsilon$ is a small number preventing singularity when $|\mathbf{x}_{ij}| = 0$ and $C_i, C_j$ respectively denote sound speeds at the $i^{th}$ and $j^{th}$ nodes.

## 3.2. Evaluation of Thermodynamic Forces

In the SPH method, the stress components are generally written in terms of hydrostatic and deviatoric stresses as,

$$\boldsymbol{\sigma}^{\alpha\beta} = -P\boldsymbol{\delta}^{\alpha\beta} + \mathbf{S}^{\alpha\beta} \tag{3.12}$$

where $P$, $\mathbf{S}^{\alpha\beta}$ are respectively the pressure and components of the traceless symmetric deviatoric stress tensor. $\boldsymbol{\delta}^{\alpha\beta}$ denotes the Kronecker delta. The pressure $P$ is usually evaluated through an equation of state (EOS). Specifically we adopt the following EOS.

$$P(\rho) = K\left(\frac{\rho}{\rho_0} - 1\right) \tag{3.13}$$

The deviatoric stress components $\mathbf{S}^{\alpha\beta}$ is evolved through the material frame indifferent Jaumann stress rate $\widehat{\dot{\mathbf{S}}}^{\alpha\beta}$, i.e. using

$$\widehat{\dot{\mathbf{S}}}^{\alpha\beta} = \dot{\mathbf{S}}^{\alpha\beta} + \mathbf{S}^{\alpha\gamma}\dot{\mathbf{R}}^{\beta\gamma} + \mathbf{S}^{\alpha\beta}\dot{\mathbf{R}}^{\alpha\gamma} \tag{3.14}$$

where
$$\dot{\mathbf{S}}^{\alpha\beta} = \mu(\overline{\dot{\mathbf{E}}}^e)^{\alpha\beta} \;;\quad \mu \text{ is the shear modulus} \tag{3.15}$$

$$(\overline{\dot{\mathbf{E}}}^e)^{\alpha\beta} = \left((\dot{\mathbf{E}})^{\alpha\beta} - \frac{1}{3}\delta^{\alpha\beta}(\dot{\mathbf{E}})^{\gamma\gamma} - (\dot{\mathbf{E}}^p)^{\alpha\beta}\right) \tag{3.16}$$

$$\dot{\mathbf{E}}^{\alpha\beta} = \frac{1}{2}\left(\frac{\partial \mathbf{v}^\alpha}{\partial \mathbf{x}^\beta} + \frac{\partial \mathbf{v}^\beta}{\partial \mathbf{x}^\alpha}\right) \tag{3.17}$$

$$\dot{\mathbf{R}}^{\alpha\beta} = \frac{1}{2}\left(\frac{\partial \mathbf{v}^\alpha}{\partial \mathbf{x}^\beta} - \frac{\partial \mathbf{v}^\beta}{\partial \mathbf{x}^\alpha}\right) \tag{3.18}$$

Computation of the thermodynamic fluxes is accomplished after writing out the following discretized forms of constitutive equations with the stress rate replaced by its objective (Jaumann) counterpart:

$$\hat{\mathbf{S}}_i^{\alpha\beta} = \mu \left[ \sum_j \frac{m_j}{\rho_j} \left\{ \frac{1}{2}(\mathbf{v}_i^\alpha - \mathbf{v}_j^\alpha) w_{ij,\beta} + \frac{1}{2}(\mathbf{v}_i^\beta - \mathbf{v}_j^\beta) w_{ij,\alpha} - \frac{1}{3}\delta^{\alpha\beta}(\mathbf{v}_i^\gamma - \mathbf{v}_j^\gamma) w_{ij,\gamma} \right\} - (\dot{\mathbf{E}}^p)_i^{\alpha\beta} \right] + \dot{\mathbf{S}}^{\alpha\gamma}\dot{\mathbf{R}}^{\beta\gamma} + \dot{\mathbf{S}}^{\gamma\beta}\dot{\mathbf{R}}^{\alpha\gamma} \quad (3.19)$$

$$\left(\dot{\boldsymbol{\xi}}^{\text{en}}\right)_i^\alpha = S_0 l_1^2 \sum_j \frac{m_j}{\rho_j} (\upsilon_i^p - \upsilon_j^p) w_{ij,\alpha} \quad (3.20)$$

$$\left(\dot{\boldsymbol{\xi}}^{\text{dis}}\right)_i^\alpha = \frac{1}{t_r^\xi} \left[ \left\{ (S_{grad})_i \left( \frac{(d^p)_i}{d_0} \right)^{m_1} l_3^2 \sum_j \frac{m_j}{\rho_j} (\upsilon_i^p - \upsilon_j^p) w_{ij,\alpha} \right\} (R_2(T))_i - \left(\boldsymbol{\xi}^{\text{dis}}\right)_i^\alpha \right] \quad (3.21)$$

$$\dot{\pi}_i^\alpha = \frac{1}{t_r^\pi} \left[ \left\{ (S_{grad})_i \left( \frac{(d^p)_i}{d_0} \right)^{m_1} \frac{\upsilon_i^p}{(d^p)_i} \right\} (R_2(T))_i - \pi_i^\alpha \right] \quad (3.22)$$

$$\dot{\mathbf{q}}_i^\alpha = \frac{1}{t_r^q} \left[ -\kappa \sum_j \frac{m_j}{\rho_j} (T_i - T_j) w_{ij,\alpha} - \mathbf{q}_i \right] \quad (3.23)$$

### 3.3. Time Integration

A standard predictor-corrector scheme - an explicit time marching method - has been used to integrate the SPH equations. The time step size is determined from the Courant-Friedrichs-Lewy (CFL) condition as,

$$\Delta t = \min_{i \in [1, N_{node}]} \left[ c_s \left( \frac{r_i}{C_i + |\mathbf{v}_i|} \right) \right] \text{ with } c_s = 0.3 \quad (3.24)$$

where $C_i$, $\mathbf{v}_i$ and $r_i$ are respectively elastic wave speed, particle velocity and smoothing length at the $i^{th}$ node.

# 4. Numerical Simulations

Our attempt in this section would be to numerically explicate the effect of micro-inertia and relaxation time parameters at the continuum level response under high strain rates. For this, impact of two 4340 steel plates fired with equal and opposite velocity is considered. A two-dimensional plane strain model (Libersky and Petscheck, 1991) is adopted for the numerical work. The initial dimensions, initial velocity and material parameters of the specimens are given in Tables 1 and 2. Each specimen is discretized with 1491 nodes - the artificial viscosity parameters being $\alpha_1 = \alpha_2 = 1$. Time step for numerical integration is chosen as 1e-7 sec.

Table 1: Material Properties (4340 steel)

| | | | |
|---|---|---|---|
| E (GPa) | 200 | $t_r^\pi$ (sec) | 9.00E-05 |
| $\nu$ | 0.29 | $t_r^\xi$ (sec) | 1.00E-06 |
| $\rho_0$ (kg/m$^3$) | 7830 | $t_r^q$ (sec) | 1.00E-07 |
| $\kappa$ (W/mK) | 38 | $m_1$ | 0.2 |
| $C_p$ (J/kgK) | 477 | $m_2$ | 2 |
| $T_r$ (K) | 298 | $n$ | 1 |
| $T_m$ (K) | 1793 | $l_1$ (mm) | 5.00E-02 |
| $d_0$ (1/s) | 1 | $l_2$ (mm) | 1.00E-05 |
| $S_0$ (MPa) | 792 | $l_3$ (mm) | 1.00E-05 |
| $H_0$ (MPa) | 510 | $l_m^2$ (mm$^2$) | 10.00 |
| $C_v$ (J/K) | 477 | | |

Table 2: Initial dimensions and velocity of specimen

| Material | $L_0$ (mm) | $D_0$ (mm) | $v_0$ (m/s) |
|---|---|---|---|
| 4340 Steel | 25.4 | 7.62 | 208 |

## 4.1. Results

In order to establish how each of the constitutive modelling parameters (viz. length scales and relaxation time parameters) affect the response, several simulations are performed varying one parameter at a time whilst keeping the rest fixed at values noted in Table 1. Additionally the micro- and macro-parts of the internal and kinetic energies are identified based on the following decomposition.

$$\rho \dot{e} = \underbrace{\boldsymbol{\sigma}:\nabla \mathbf{v} + \rho h - \nabla \cdot \mathbf{q}}_{\text{macro-part}} + \underbrace{(\pi - \tau)\upsilon^p + \boldsymbol{\xi} \cdot \nabla \upsilon^p}_{\text{micro-part}}$$

$$\text{Kinetic power} = \underbrace{\rho \dot{\mathbf{v}} \cdot \mathbf{v}}_{\text{macro-part}} + \underbrace{\rho l_m^2 \dot{\upsilon}^p \upsilon^p}_{\text{micro-part}}$$

Of specific interest in the numerical study would be the plots of the final (post-impact) equivalent plastic strain profile, final temperature profile, evolutions of macro and micro parts of energy, evolution of the total energy and evolution of the equivalent plastic strain at the mid-point of the impacting surface. Our observations based on these plots are furnished below.

### 4.1.1. Length Scale $(l_m)$ Associated with Micro-Structural Defect Mass

Plastic strain and temperature profiles shown in Figures 4.1 and 4.2 respectively reflect the effect of a varying $l_m^2$ on these quantities. Specifically, the decreasing value as well as a lesser spread of the equivalent plastic strain and temperature with increasing $l_m^2$ indicate higher and more localized resistance to plastic flow with the increasing fictive mass of micro-structural defects.

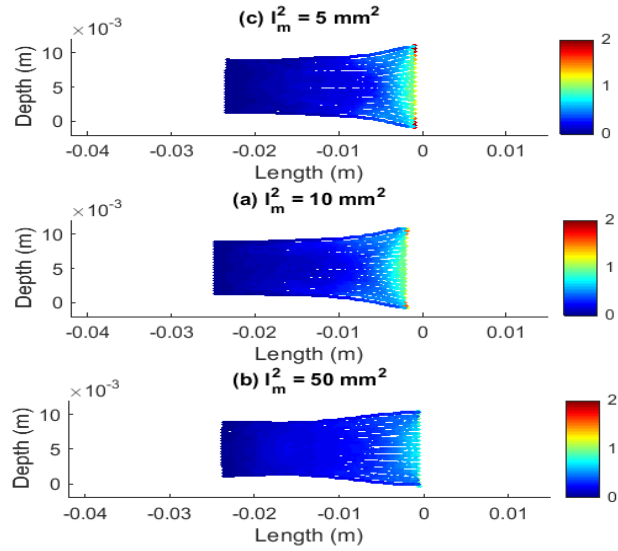

Figure 4.1. Effect of $l_m$ on the equivalent plastic strain profile

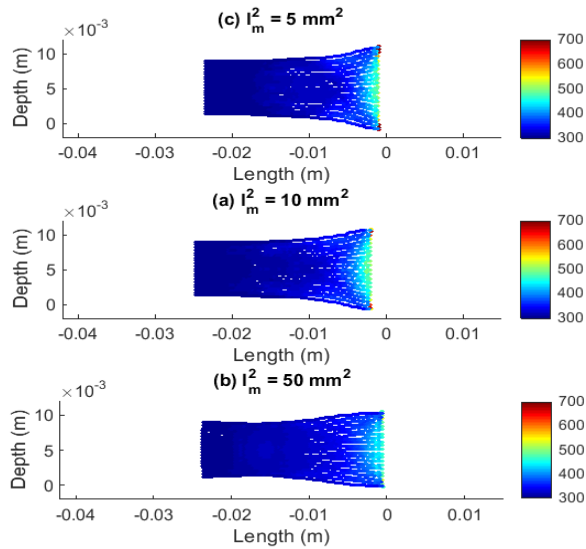

Figure 4.2. Effect of $l_m$ on the temperature profile

The effect of micro-structural defect mass on the kinetic energy (Kin), internal energy (Int) and total energy (Total) is shown in Figures 4.3, 4.4 and 4.5. A consistent increase in the micro-part of the kinetic and internal energies with increasing $l_m$ may significantly affect the respective macro-parts. Micro-inertial effects are indeed very pronounced in the plot exhibiting the evolving equivalent plastic strain at the mid-point of the impact surface and a stiffening against plastic deformation with increasing $l_m$ is observed; see Figure 4.6.

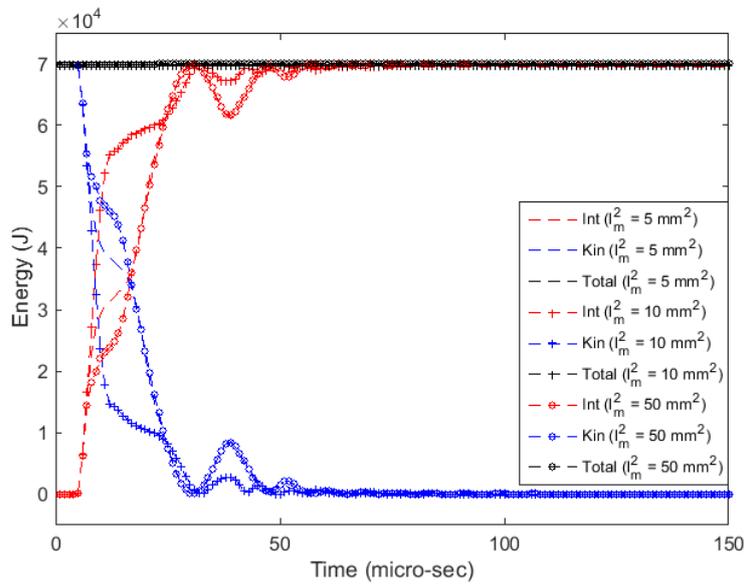

Figure 4.3. Effect of $l_m$ on macro-parts of energy

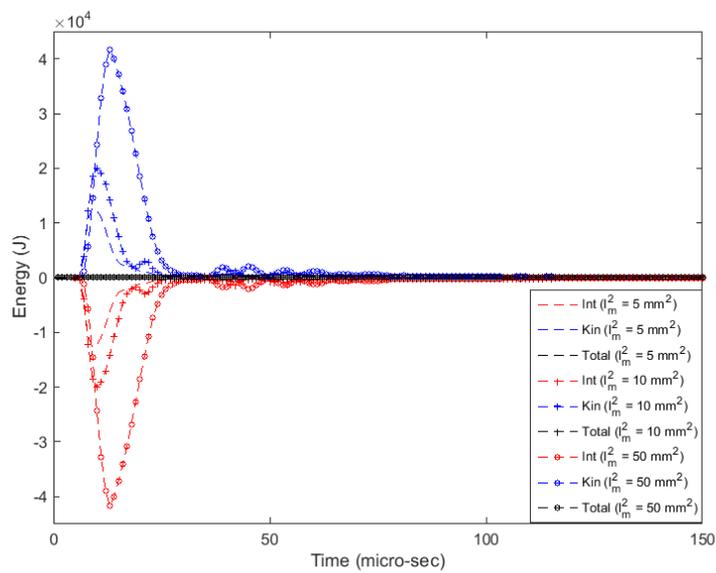

Figure 4.4. Effect of $l_m$ on micro-parts of energy

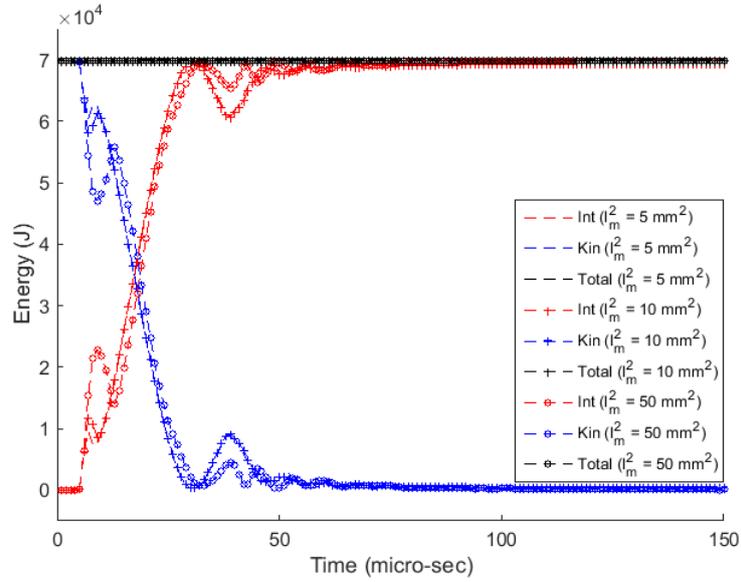

Figure 4.5. Effect of $l_m$ on kinetic and internal energy (each a sum of micro- and macro-parts)

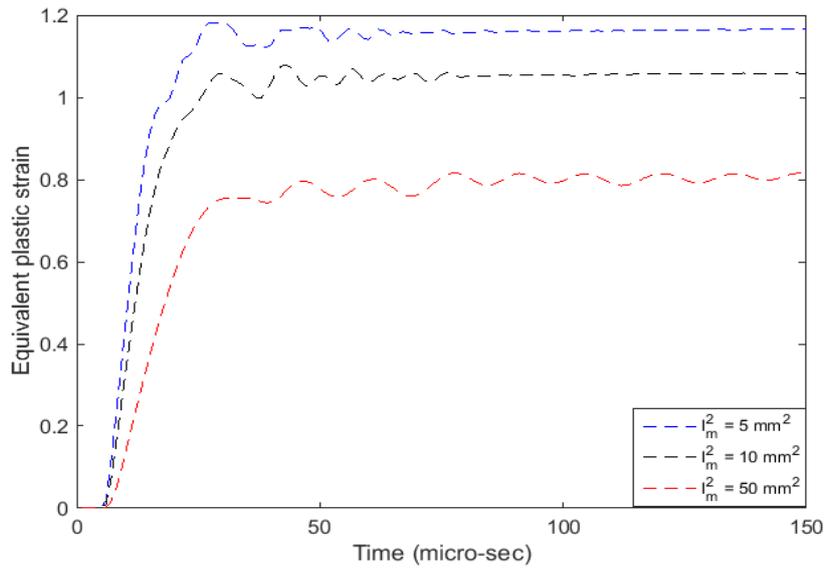

Figure 4.6. Effect of $l_m$ on evolution of equivalent plastic strain at mid-point of impact surface

### 4.1.2. Relaxation Time Parameters

We now demonstrate how the micro-stresses and the heat flux depend on the relaxation time parameters. Typically, the kinematic restraint associated with the gradient of the plastic strain is more than the plastic strain itself. Accordingly, the relaxation time parameter $t_r^\xi$ associated with the vector micro-stress (which is conjugate to the gradient of the plastic strain rate) should be at least an order smaller than the relaxation time $t_r^\pi$ for the micro-stress (conjugate to the plastic strain rate). Moreover, the relaxation time parameter $t_r^q$ associated with the heat flux is in general very small for materials. Figures 4.7 and 4.8 show that, as the relaxation time $t_r^\pi$ increases, so do the plastic flow and the associated temperature. This observation is consistent with the fact that materials having higher relaxation times tend to be more visco-plastic.

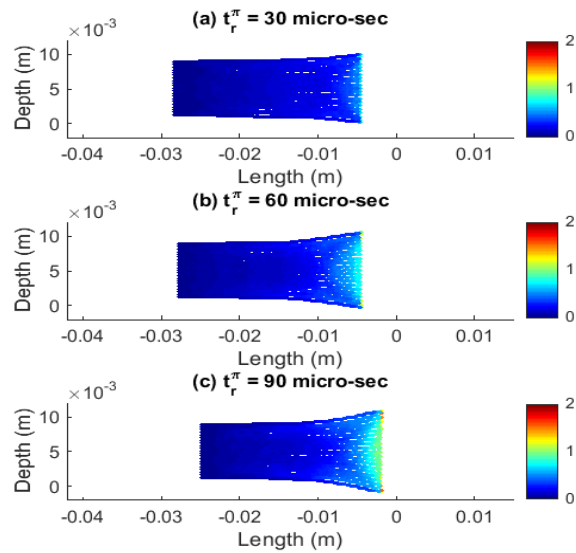

Figure 4.7. Effect of $t_r^\pi$ on equivalent plastic strain profile

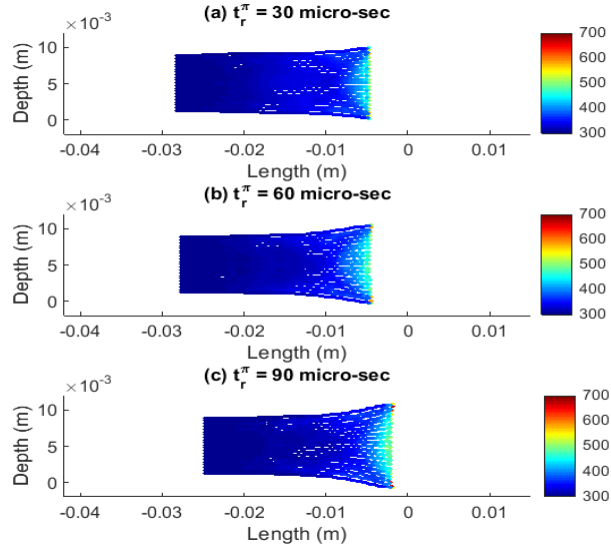

Figure 4.8. Effect of $t_r^\pi$ on temperature profile

Figures 4.9 through 4.12 are suggestive of a low sensitivity of the plastic response and the heat flux to variations in $t_r^\xi$, the relaxation time associated with the vector micro-stress. The gradient theory could be thought as a constrained Cosserat therory, wherein $\gamma^p$ and $\nabla \gamma^p$ could be respectively associated with the plastic slip and a rotational component of motion that inevitably arises to enforce displacement continuity (see, for instance, Fleck *et al.*, 1994; Section 3). Recalling that $\xi$ is the thermodynamic conjugate force to $\nabla \gamma^p$ and that micro-structural rotation relaxes much faster than translation, one indeed anticipates less sensitivity of the plastic response to changes in $t_r^\xi$.

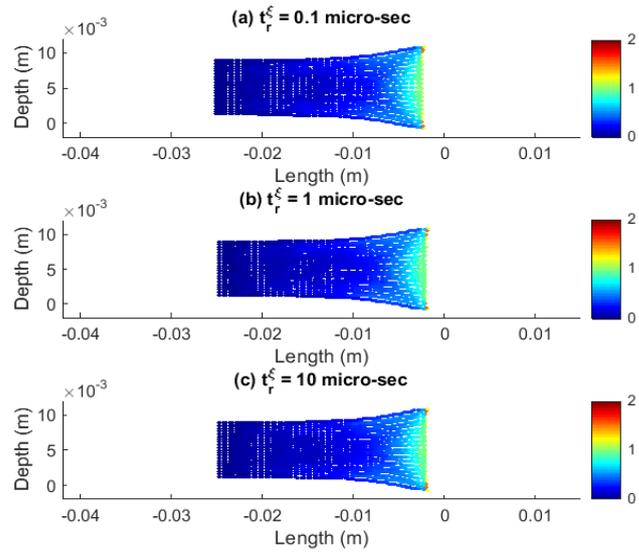

Figure 4.9. Effect of $t_r^\xi$ on equivalent plastic strain profile

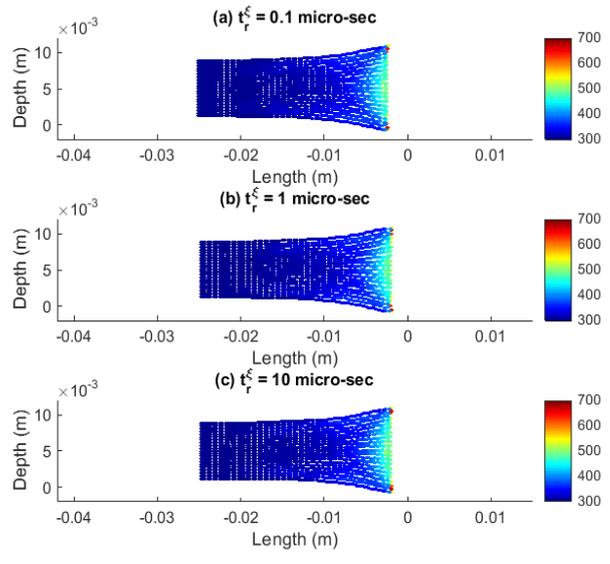

Figure 4.10. Effect of $t_r^\xi$ on temperature profile

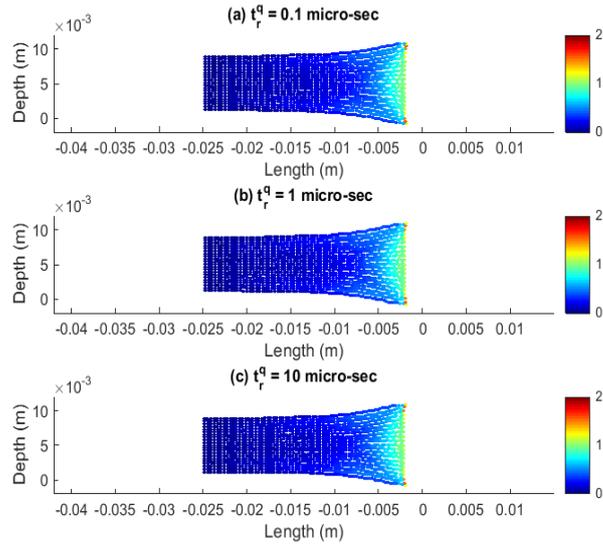

Figure 4.11. Effect of $t_r^q$ on equivalent plastic strain profile

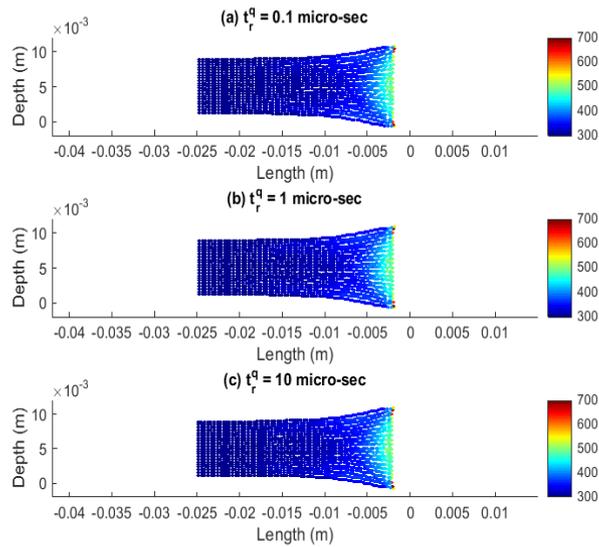

Figure 4.12. Effect of $t_r^q$ on temperature profile

Variations in energy with changes in the relaxation time parameters associated with the dissipative fluxes (i.e., $\pi$, $\xi^{\text{dis}}$ and $\mathbf{q}$) are presented in Figures 4.13 - 4.21. As $t_r^\pi$ increases, so does the contribution from the micro-part of energy (Figure 4.13) and the time-dependent interaction between the total (as well as the macro-components of) kinetic and internal energy takes place over a longer time (see Figures 4.14 and 4.15). In contrast, changes in $t_r^\xi$ and $t_r^q$ do not meaningfully contribute to alter the energy (Figures 4.16 through 4.17).

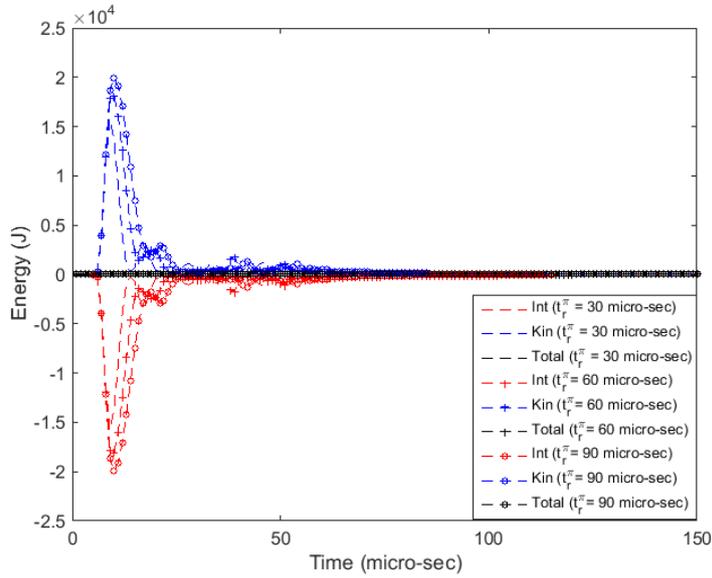

Figure 4.13. Effect of $t_r^\pi$ on micro-parts of energy

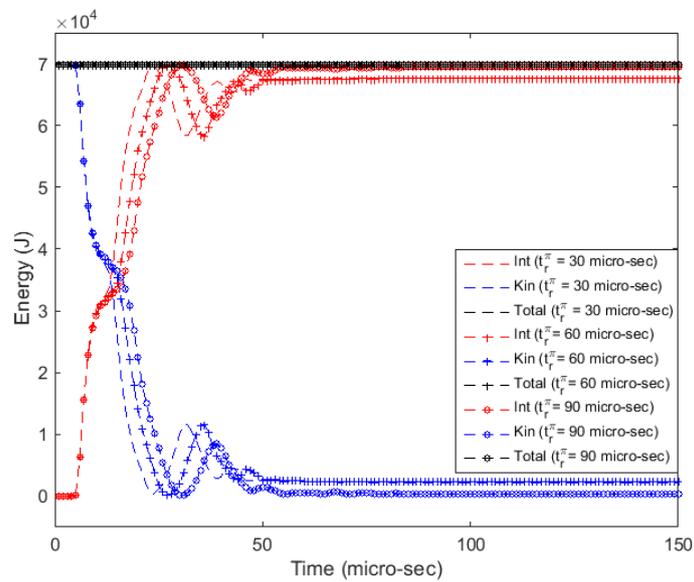

Figure 4.14. Effect of $t_r^\pi$ on macro-parts of energy

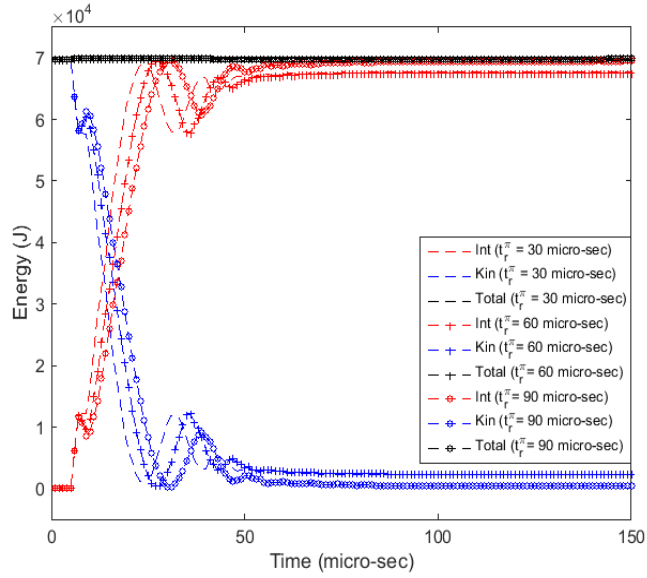

Figure 4.15. Effect of $t_r^\pi$ on kinetic and internal energy (each a sum of micro- and macro-parts)

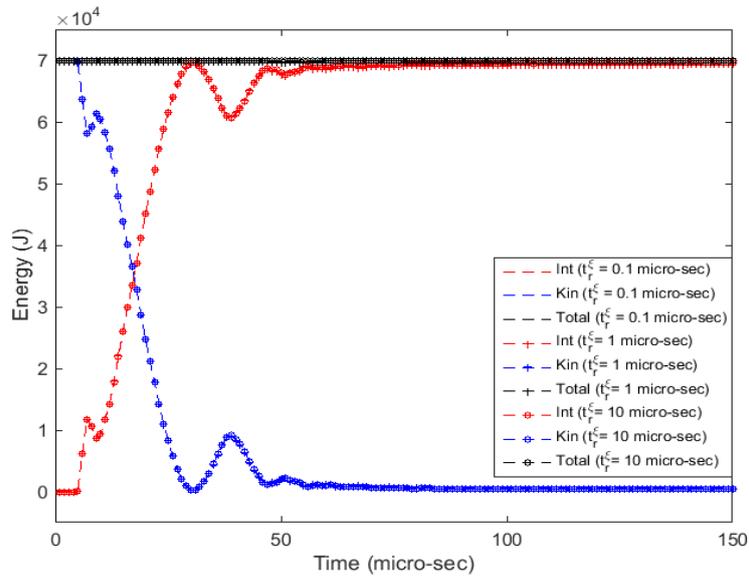

Figure 4.16. Effect of $t_r^\xi$ on kinetic and internal energy (each a sum of micro- and macro-parts)

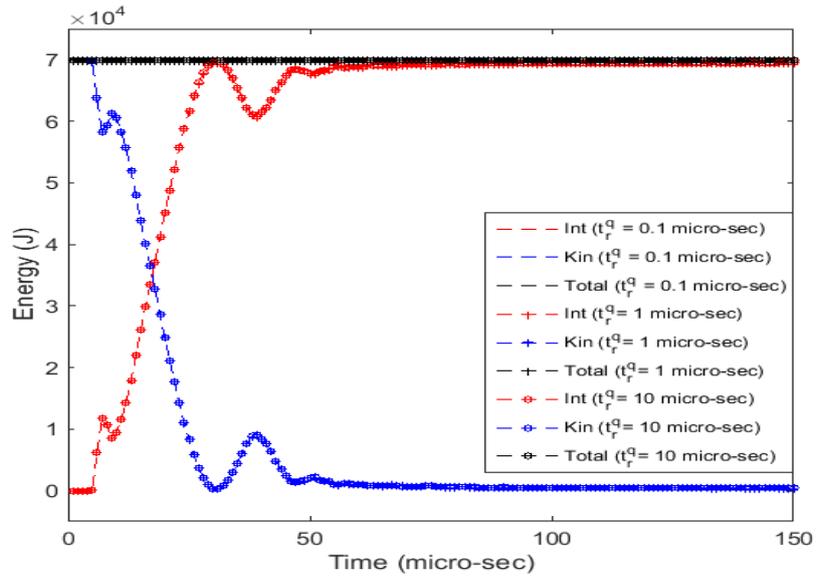

Figure 4.17. Effect of $t_r^q$ on kinetic and internal energy (each a sum of micro- and macro-

parts)

As the effect of $t_r^\pi$ on the overall response is quite pronounced, our interest is also on assessing how it affects the plastic strain as it evolves. Indeed, significant changes in the plastic strain evolution, again sampled at the mid-point of the impact surface, may be observed as $t_r^\pi$ varies (Figure 4.18).

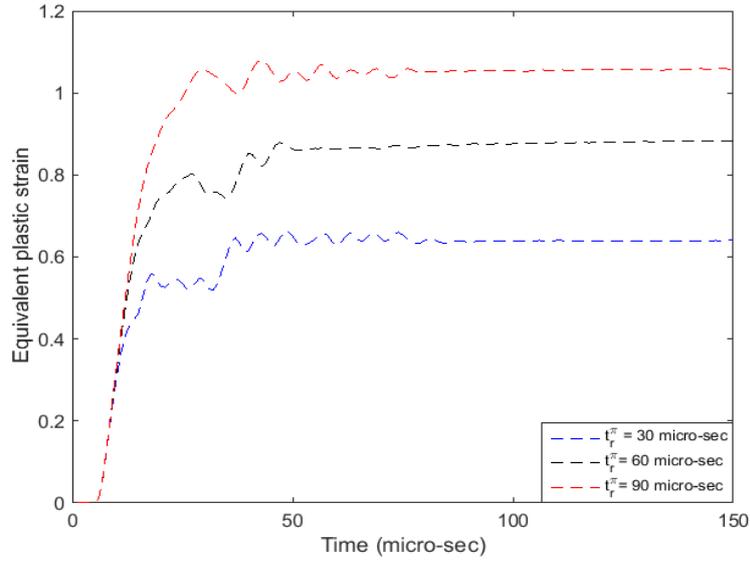

Figure 4.18. Effect of $t_r^\pi$ on evolution of equivalent plastic strain at mid-point of impact surface

### 4.1.3. Energetic and Dissipative Length Scales

It is well known (Voyiadjis and Faghihi, 2012) that the energetic length scale ($l_1$) and the dissipative length scales ($l_2$ and $l_3$) are introduced to account for size effects. In line with this, we wish to establish the sensitivity of these length scales to the overall impact dynamic response. Figures 4.19 through 4.24 show that these have quite a negligible influence on the equivalent plastic strain and temperature profiles. This is not unexpected, especially given that the dynamic response is dominated by inertial effects that are in turn controlled by both the mass density and relaxation time parameters.

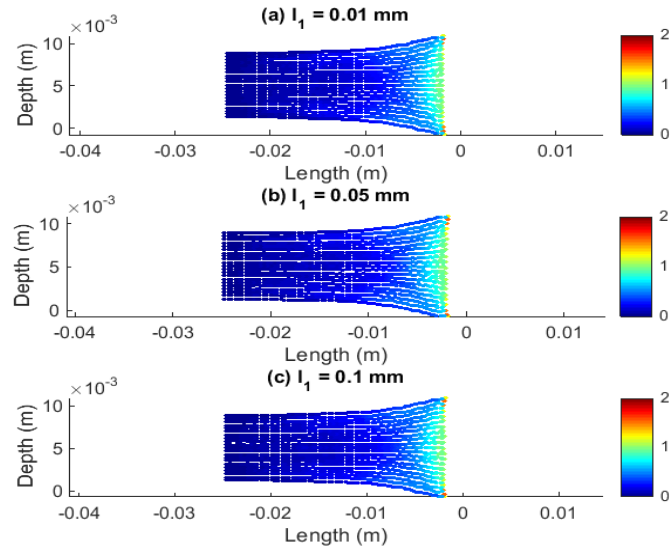

Figure 4.19. Effect of $l_1$ on equivalent plastic strain profile

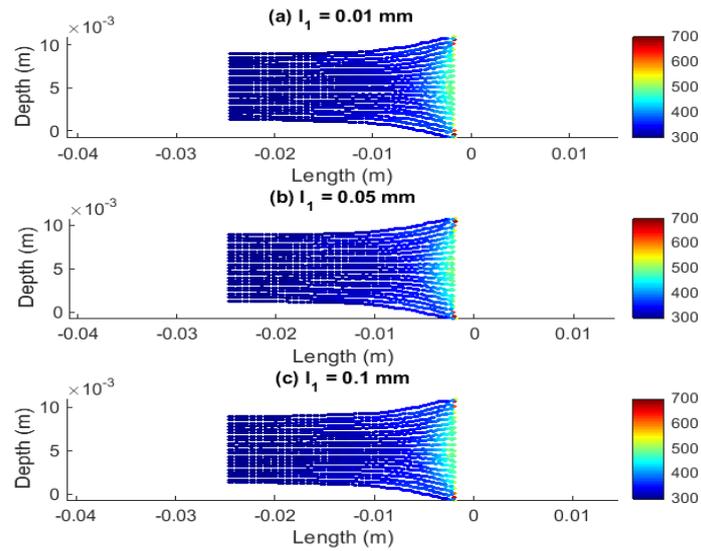

Figure 4.20. Effect of $l_1$ on temperature profile

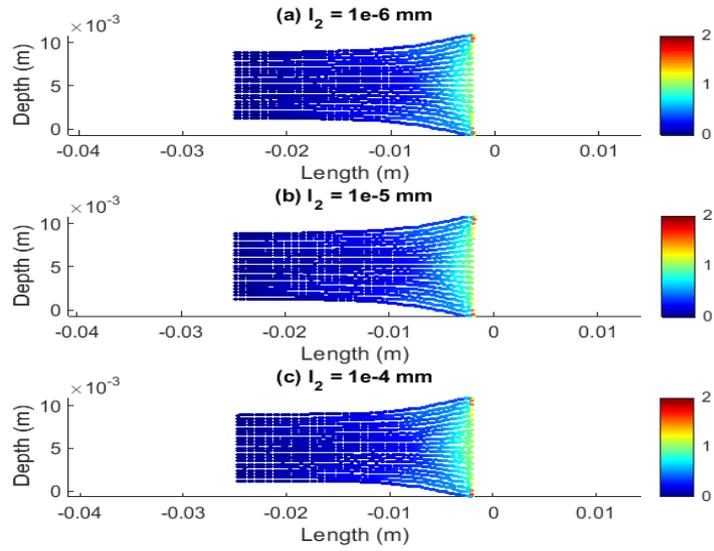

Figure 4.21. Effect of $l_2$ on equivalent plastic strain profile

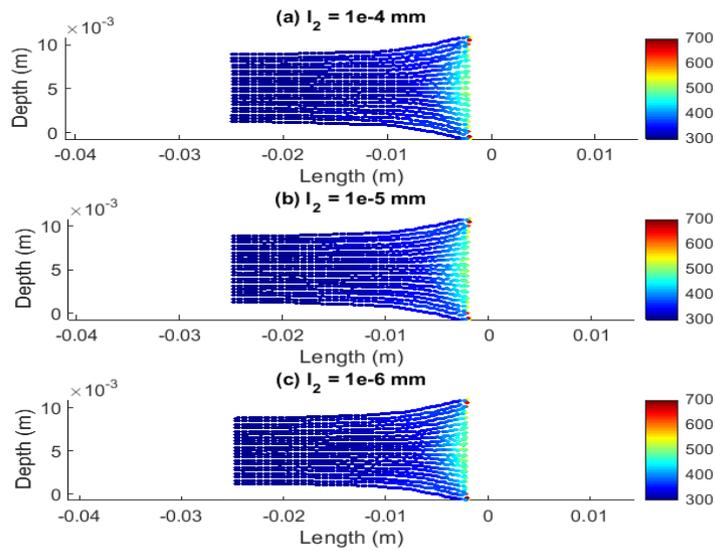

Figure 4.22. Effect of $l_2$ on temperature profile

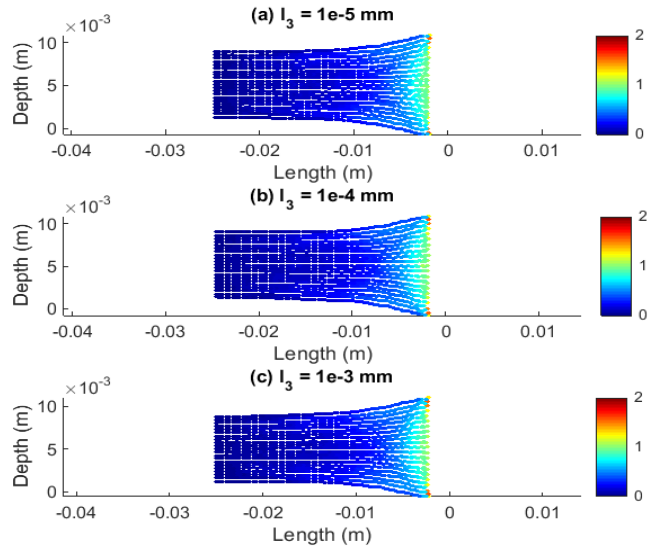

Figure 4.23. Effect of $l_3$ on equivalent plastic strain profile

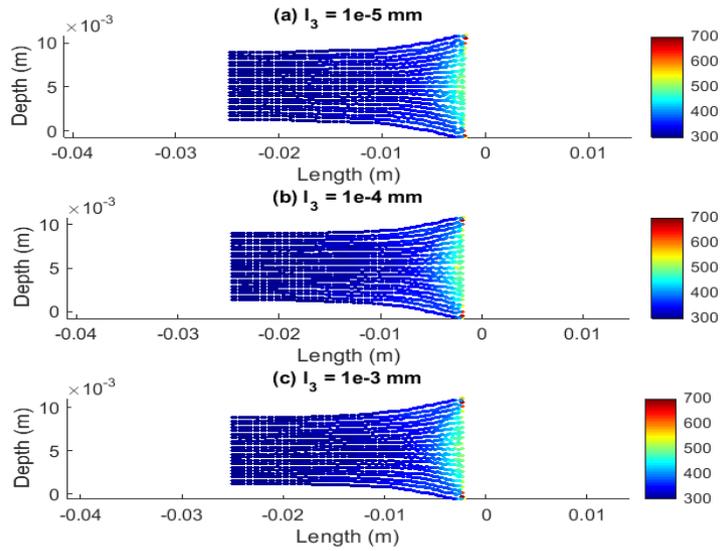

Figure 4.24. Effect of $l_3$ on temperature profile

Figures 4.25 to 4.27 are presented to show that the length scale parameters $l_1$, $l_2$ and $l_3$ hardly affect the energy evolutions.

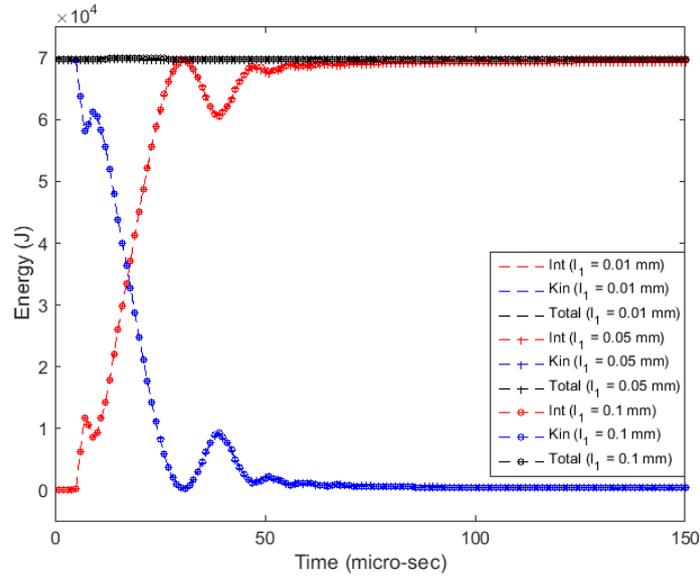

Figure 4.25. Effect of $l_1$ on kinetic and internal energy (each a sum of micro- and macro-parts)

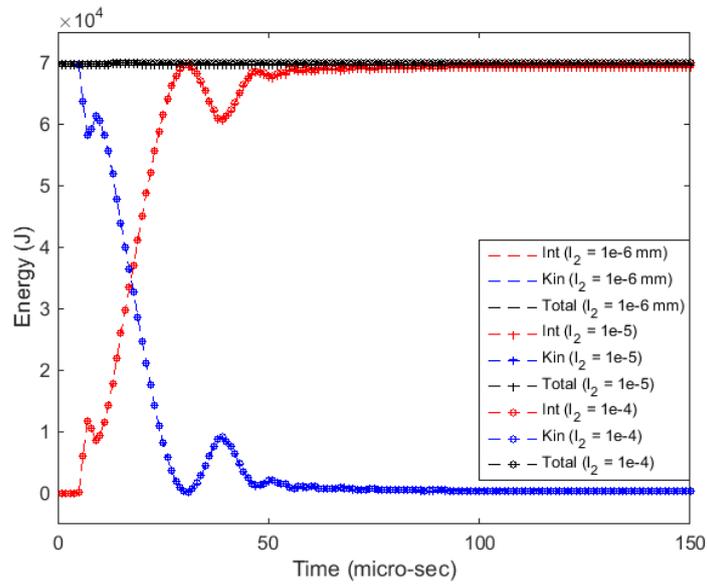

Figure 4.26. Effect of $l_2$ on kinetic and internal energy (each a sum of micro- and macro-parts)

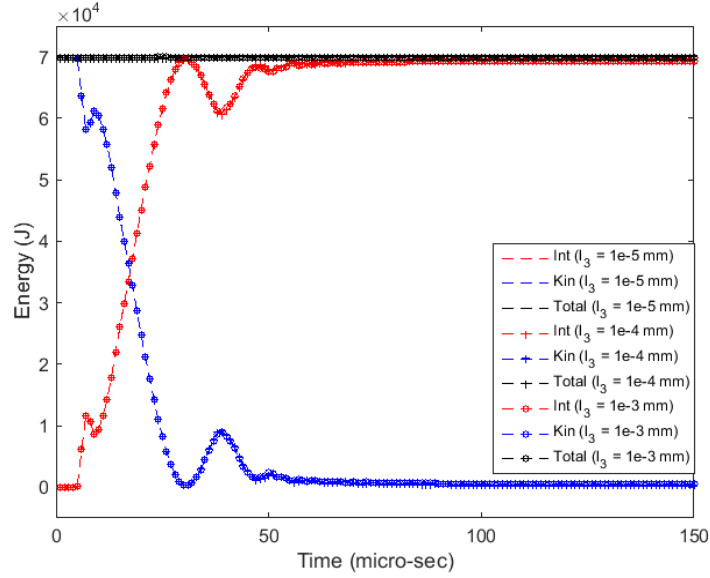

Figure 4.27. Effect of $l_3$ on kinetic and internal energy (each a sum of micro- and macro- parts)

Finally, with $T$ denoting a time within which the post-impact response reaches a steady-state, we show in Figure 4.28 a plot of the so called maximal flow stress $\tau_{max} = \sup\{\tau(t): t \in [0, T]\}$ as a function of the impact velocity. In view of the fact that a higher impact velocity implies a higher maximum value of the achieved plastic strain rate, the plotted curve closely resembles the well-known pattern of an anomalous increase in hardening for strain rates higher than $\sim 10^4 \text{ s}^{-1}$, at least when the plastic strain is small (Follansbee and Kocks, 1988; Preston *et al.*, 2003). Interpretation of the numerical results presented herein must only be to facilitate a qualitative understanding until a material-specific quantitative assessment is made possible upon the determination of the time and length scale parameters through experiments. Pending the latter exercise currently under way, the values assigned to these parameters are chosen somewhat arbitrarily even though the general trends reported should not depend on the choice.

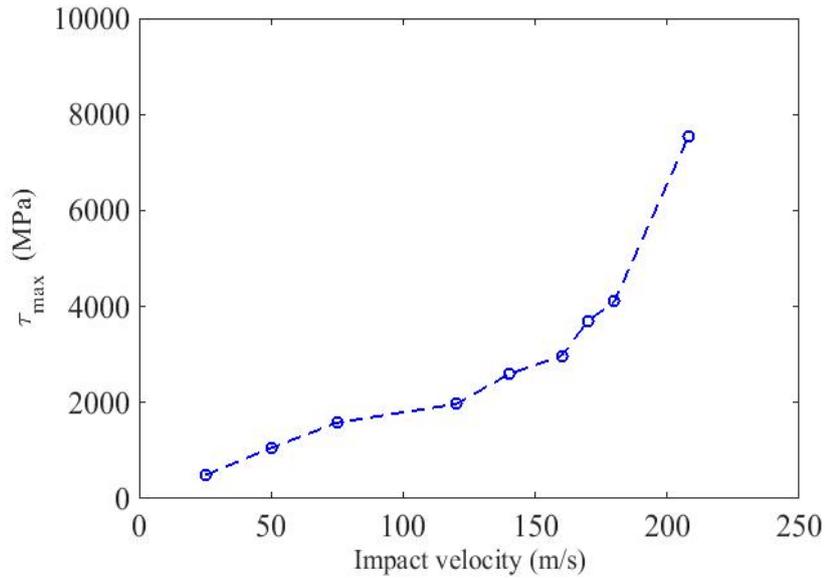

Figure 4.28. Variation of $\tau_{max}$ with increase in impact velocity

# 5. Conclusions

A micro-inertia driven dynamic flow rule, also referred to as the micro-force balance that appears as an initial-boundary value problem, is derived consistent with the underlying principles of rational thermodynamics. It is employed within a thermo-visco-plasticity theory of the continuum with a view to modelling the response of isotropic solids under high strain rates. The dynamic flow rule evolves the equivalent plastic strain. For a given dislocation density in polycrystalline solids, the plastic strain rate is considered an averaged macroscopic manifestation of the dislocation velocity that in turn multiplies with a length-scale dependent fictive mass of the moving line defects to yield the micro-inertia term. Relaxation time parameters are introduced in the constitutive modelling of the dissipative fluxes leading to Maxwell-Cattaneo type equations for their evolutions, thus resolving the paradox of infinite speed of propagation of disturbances. As anticipated, the reported numerical simulations are suggestive of an impeded plastic motion as the fictive mass of the moving defects increases. In addition, only the micro-inertial length scale and the relaxation time for the scalar micro-stress appear to substantively influence the evolving energy landscape as it is split into the micro- and macro-components of the internal and kinetic energies.

Plastic response, quasi-static or dynamic, is essentially a non-equilibrium thermodynamic phenomenon that is hardly expected to conform to the principles of rational thermodynamics exploited throughout this work. Despite inserting *caveat lectors* at the appropriate places, time appears ripe to exploit some of the integral fluctuation relations in stochastic thermodynamics in lieu of the conventional second law while developing the constitutive models in dynamic plasticity.